# How Does Folding Modulate Thermal Conductivity of Graphene?

Nuo Yang[1,2], Xiaoxi Ni[2],Jin-Wu Jiang[2], Baowen Li[2, 1, #]

[1]Center for Phononics and Thermal Energy Science, Department of Physics, Tongji University, 200092 Shanghai, China,

[2]Department of Physics and Centre for Computational Science and Engineering, National University of Singapore, Singapore 117542, Republic of Singapore

[#] To whom correspondence should be addressed. E-mail: (B.L.) phylibw@nus.edu.sg

**Abstract**

We study thermal transport in folded graphene nanoribbons using molecular dynamics simulations and the non-equilibrium Green's function method. It is found that the thermal conductivity of flat graphene nanoribbons can be modulated by folding and changing interlayer couplings. The analysis of transmission reveals that the reduction of thermal conductivity is due to scattering of low frequency phonons by the folds. Our results suggest that folding can be utilized in the modulation of thermal transport properties in graphene and other two dimensional materials.

Graphene, one-atom-thick planar sheets of sp2-bonded carbon atoms that are densely packed in a honeycomb crystal lattice, has recently become the focus of scientific community due to its unique electronic, phononic and optical properties, and its potential to become the mainstream semiconductor materials in future device fabrications.[1] The thermal properties of graphene are important both for fundamental understanding of the underlying physics in low-dimensional system and for applications. Superior thermal conductivity has been observed in graphene,[1,2] which has raised the exciting prospect of using graphene structures for thermal phononic devices.

Graphene nanoribbons (GNR), which are patterned as thin strips of graphene, are also known to display diverse transport properties[3-5] compared with infinite sheet, due to the possibility of manipulating different ribbon width as well as the atomic configuration of the edges in the GNRs. Recently, it is demonstrated that thermal rectification effect in the asymmetric GNRs.[6-8] The dependents of thermal conductivity of GNR on the size were studied by equilibrium molecular dynamics (EMD)[9] and non-equilibrium MD (NEMD) simulation.[10] The stress/strain effects on the thermal conductivity of low-dimensional silicon and carbon materials[11] and graphene ribbons[12] were studied by EMD simulations.

In 2008, the 1+1 folded graphene has been successfully fabricated and their electronic structures have been investigated by Raman spectroscopy.[13] It is found that the folded graphene have different electronic structures compared to monolayer graphene, which opens an extra room to manipulate the properties of graphene derivatives - folding. Recently, the multiply folded graphene, termed *grafold*, has been realized experimentally.[14,15] The scientists utilize folding as a technique to alter the properties of graphene, but limited to the electronic transport regime.

However, it is still not clear how and why the folding can alter the thermal properties. In this letter, we will answer these questions.

The structures of folded GNRs are shown in Fig. 1(a)-(h). The structures are zigzag GNRs with width 0.71 nm (4 atoms in each layer). A few layers closed to the two boundaries of GNRs are treated as heat source and sink respectively. The length of GNR between the heat source and sink is same for each case, which is set as 21.3 nm (173 layers). In order to compare the effect of folds, the completely flat zigzag GNR with the same size is also studied. The two ends of GNRs are applied with the fixed boundary conditions. To mimic the graphene is placed on top of a substrate in experiments, we applied the substrate coupling effect below the bottom layer of each multiply folded GNR (shown in Fig.1 (b), (d), (f), and (h)), and the single layer flat GNR. The couplings between GNR and the substrate are van der Waals interactions which are modeled by the Lennard-Jones (LJ) potential as

$$V_{ij} = 4\chi\varepsilon_{ij}\left[\left(\frac{\sigma_{ij}}{r}\right)^{12} - \left(\frac{\sigma_{ij}}{r}\right)^{6}\right] \qquad (1)$$

where $\chi$ is a scaling factor, $\varepsilon$ is the energy parameter, $\sigma$ is the distance parameter and r is the interatomic distance. For the parameterization of the potential, the reader is referred to the earlier work.[16]

To observe the compressing effect on thermal transport in multiply folded GNRs, another substrate coupling is put on the top layer of each multiply folded GNR (shown in Fig.1 (b), (d), (f), and (h)). The compressing processes are done during the relaxation of the structure. In the beginning of relaxation, the interlamellar space is 0.74 nm (shown in Fig.1 (a), (c), (e), and (g)). Following, we move the positions of the top substrate in each folded structures a small distance toward the bottom layer at every first $10^{6}$ time steps. This moving distance of each step ($\sim 10^{-6}$ nm) is small enough comparing the movement of atoms at each time step (~$10^{-4}$ nm) to avoid destroying the curvature structure at the end of the compressing process. After applying different movement distances of the top substrate, some relaxed structures of multiply folded graphene nanoribbons are displayed in the right channel of Fig.1 (Fig. 1 (b), (d), (f) and (h)).

All the multiply structures are optimized under the second generation reactive empirical bond order Brenner potential[17] and NEMD simulation is implemented. The atoms in a few layers close to the two ends of GNRs are placed in the Nosé-Hoover thermostats[18,19] at temperature $T_{top} = (1 + \Delta) \cdot T$ and $T_{bot} = (1 - \Delta) \cdot T$, T and $\Delta$ is set as 300 K and 0.1, respectively. The equations of motion are integrated by the velocity Verlet algorithm.

The time step is set as 0.5 fs, and the simulation runs for $10^7$ time steps giving a gross simulation time of 5 ns. Simulations are carried out long enough (the first 1ns) such that the system reaches a steady state. Then, the kinetic temperature and heat flux are averaged from 1 ns to 5 ns. Since the convergence time is two times longer than the relaxation time, its choice is independent of the structures under consideration. The heat current is recorded from the power of heat bath. The total heat flux is defined as

$$J = \frac{1}{N_t} \sum_{i=1}^{N_t} \frac{\Delta \varepsilon_i}{2 \Delta t} \qquad (2)$$

where $\Delta \varepsilon$ is energy added/removed out at each heat bath for each time step $\Delta$t. The thermal conductivity κ can be calculated from Fourier's law,

$$\kappa = -\frac{J \cdot L}{A \times (T_{top} - T_{bot})} \qquad (3)$$

where A is the cross section area, and L is the length of the folded GNR from the heat source to sink.

The advantage of MD method is that it does not rely on any thermodynamic limit assumptions and is thus applicable for any system size in principle, which is important for the investigation of realistic folding nanoscale systems in this study. To overcome the drawback of purely classical property of MD simulation, the temperature calculated in the classical MD simulation ($T_{MD}$) has been corrected by taking into account the quantum effects of phonon occupation[8]

$$T_{MD} = \frac{2T^3}{T_D^2} \int_0^{T_D/T} \frac{x^2}{e^x - 1} dx \qquad (4)$$

where $T_D$ and T are the Debye temperature and corrected temperature respectively. $T_{MD}$ is 207 K for T=300 K.

Fig. 2 shows the modulation of thermal conductivity for GNRs with different numbers of folds. The value of relative thermal conductivity of 1 corresponds to 111.51 Wm$^{-1}$K$^{-1}$, which is the thermal conductivity of a flat GNR with substrate. When the interlamellar space is 0.74 nm, the thermal conductivity is reduced to 71% by 3 folds in GNR. For the same interlamellar space, the thermal conductivity is decreased to 49% in GNR with 6 folds. The results show the more numbers of folds are there, the lower of the thermal conductivity due to the scatterings of phonons at the folds.

Meanwhile, it is also displayed in Fig. 2 that modulation of the thermal conductivity by compressing interlamellar space. Generally, the more one compresses the distance between the first and last layer, the smaller of the value of thermal conductivity. For example, in GNR with 3 folds, the thermal conductivity is reduced to 59% by compressing interlamellar space to 0.24 nm. The phenomenon of reduction becomes more obvious in structures with more folds. For instance, the thermal conductivity of GNR with 6 folds can be reduced to 31% by compressing interlamellar space to 0.27 nm. The van der Waals interactions cause the interlayer scatterings. As the interlamellar space decrease, the interactions and scatterings increase which is shown in details the following section. However, one can also discover a different trend of thermal conductivity alteration, for example, in the structure with 5 folds, thermal conductivity is enhanced when the interlamellar space is decreased from 0.64 nm to 0.44 nm. Nevertheless, the thermal conductivity can be reduced further when the interlamellar space is compressed more.

It has been predicted ZA modes contribute up to 75% in the contribution of thermal conductivity in graphene at room temperature.[20] Our calculation shows that the folds have large effect on the ZA modes in GNR. As shown in Fig. 3(a), the ZA modes in the GNR with folds are combination of out-of-plane and in-plane modes, instead of just out-of-plane modes in graphene sheet. When the phonon transfer along the folded GNR from one end to the other, the ZA modes need change from out-of-plane into in-plane modes at each fold, and change back to in-plan modes after passing the folds.

The ballistic Non-Equilibrium Green's Function (NEGF) approach is employed to calculate the phonon transmission for both a flat GNR and a folded GNR with infinite length. In the system, the center region (e.g. the flat GNR or the folded GNR) is sandwiched by two leads (quantum heat baths). The phonon transmission coefficient can be obtained by using the Caroli formula:

$$T[\omega] = Tr\left(G^r \Gamma_L G^a \Gamma_R\right)$$

$$G^a = \left(G^r\right)^+, \Gamma_L = -2\,\mathrm{Im}\left(\Sigma_L\right)$$

(5)

where $G^r$ is the retarded Green's function for the center region, $\Gamma_L(\Gamma_R)$ is the self-energy for the left/right lead (the details of NEGF method are in Ref. 21). To show the scattering effect of the folds on the GNR, the spectra of transmission ratio, $T_R[\omega] = T_{folded}[\omega]/T_{flat}[\omega]$, is shown in Fig. 3(b) and (c). Generally, the high frequency phonons are easily scattered by impurities, dislocation, and boundaries etc..[22] Unlike the other scatterings, the low frequency transmission is largely decreased by folds, instead of high frequency.

Our MD simulation results show the thermal conductivity is decreased by not only folding GNR and also compressing interlamellar space. It is shown in Fig. 4 that the phonon dispersion curves of graphite calculated by General Utility Lattice Program (GULP).[23] Fig. 4 (a) and (b), it shows the dispersion curve of graphite for the interlamellar space at equilibrium (0.337 nm) and compressed (0.27 nm), respectively. The high conductivity of graphene sheet comes from the limitation of symmetry-based phonon scattering selection rules which state as only even numbers of out-of-plane phonons (ZA and ZO) can be involved in anharmonic phonon-phonon scattering. The graphite and folded GNR have lower symmetry than graphene ($D_{6h}$), and then the anharmonic scatterings dominate the thermal conductivity of the folded GNR at high temperature. Compressing interlamellar space could increase the anharmonic scatterings in GNR. In Fig. 4(c) and (d), it is shown that there are much more number of phonon states available for three-phonon Umklapp scatterings, because the dispersion of ZA mode changes from a quadratic to linear, based on Van de walls Forces interactions among graphene layers.

In conclusion, we have demonstrated that the folding is a useful method to modulate the thermal transport properties in GNRs. Our MD simulation results shows the thermal conductivity can be substantially decreased up to 70% compared to its flat analogue. The percentage of reduction is dependent on the number of folds. Moreover, the more the structure with folds is compressed, the more the thermal conductivity is reduced. The transmission spectra calculated by NEGF show the decrease of thermal conductivity comes from strong scattering of low frequency modes at the folds. In addition, compressing the interlamellar space provide more number of phonon states available for three-phonon scatterings. Our results suggest a promising way for future phonon engineering in graphene derivatives. Besides geometries, sizes and other modulation methods, the method of folding provides additional freedom of manipulation transport properties in graphene, which may further strengthen its position as a mainstream building block in future device fabrication.

NY and XN contributed equally to this work. This work is supported in part by the startup fund from Tongji University, China (NY). NY wishes to acknowledge useful discussions with Jie Chen and Lifa Zhang.

**Legends**

Fig. 1: Simulation structures of multiply folded GNRs. From (a) to (b), the structures are GNRs before relaxation and have 0.74 nm in interlamellar space. (a) GNR with 3 folds; (b) Relaxed structure of GNR with 3 folds and 0.24 nm interlamellar space; (c) GNR with 4 folds; (d) Relaxed structure of GNR with 4 folds and 0.37 nm interlamellar space; (e) GNR with 5 folds; (f) Relaxed structure of GNR with 5 folds and 0.34 nm interlamellar space; (g) GNR with 6 folds; (h) Relaxed structure of GNR with 6 folds and 0.27 nm interlamellar space. The heat bath with high temperature ($T_{top}$) and low temperature ($T_{bot}$) is show in red and blue, respectively. There are two substrates for each GNR. One is on the top layer and the other is below the bottom layer.

Fig. 2: (color online) Relative thermal conductivity modulation by compressing interlamellar space with different folds in GNRs. The value 1.0 of relative thermal conductivity corresponds to 111.5 $Wm^{-1}K^{-1}$ which is the thermal conductivity of the flat zigzag GNR. The sizes of all GNRs are the same, which are 0.71 nm in width and 21.3 nm in length.

Fig. 3: (color online) (a) A ZA mode ($\omega=50$ $cm^{-1}$) in the folded GNR with periodic boundary condition in z-axis. The ZA mode is a combination of out-of-plane mode and in-plane mode. Arrows correspond to eigenvectors. (b) The transmission ratio at a given frequency is the transmission coefficient of the folded GNR over the transmission coefficient of the flat GNR. It shows the scattering effect on the GNR by the folds. (c) The spectra of phonon transmission coefficient of the flat GNR and the folded GNR calculated by NEGF.

Fig. 4: (a) and (b) The phonon dispersion curve of graphite (AB Bernal stacking) without and with compressing. The interlamellar space is 0.337 nm and 0.27 nm, respectively. (c) and (d) Diagram of three-phonon Umklapp scattering in graphite without and with compressing, corresponding to (a) and (b), respectively. It shows that there are much more number of phonon states available for scatterings when compressing.

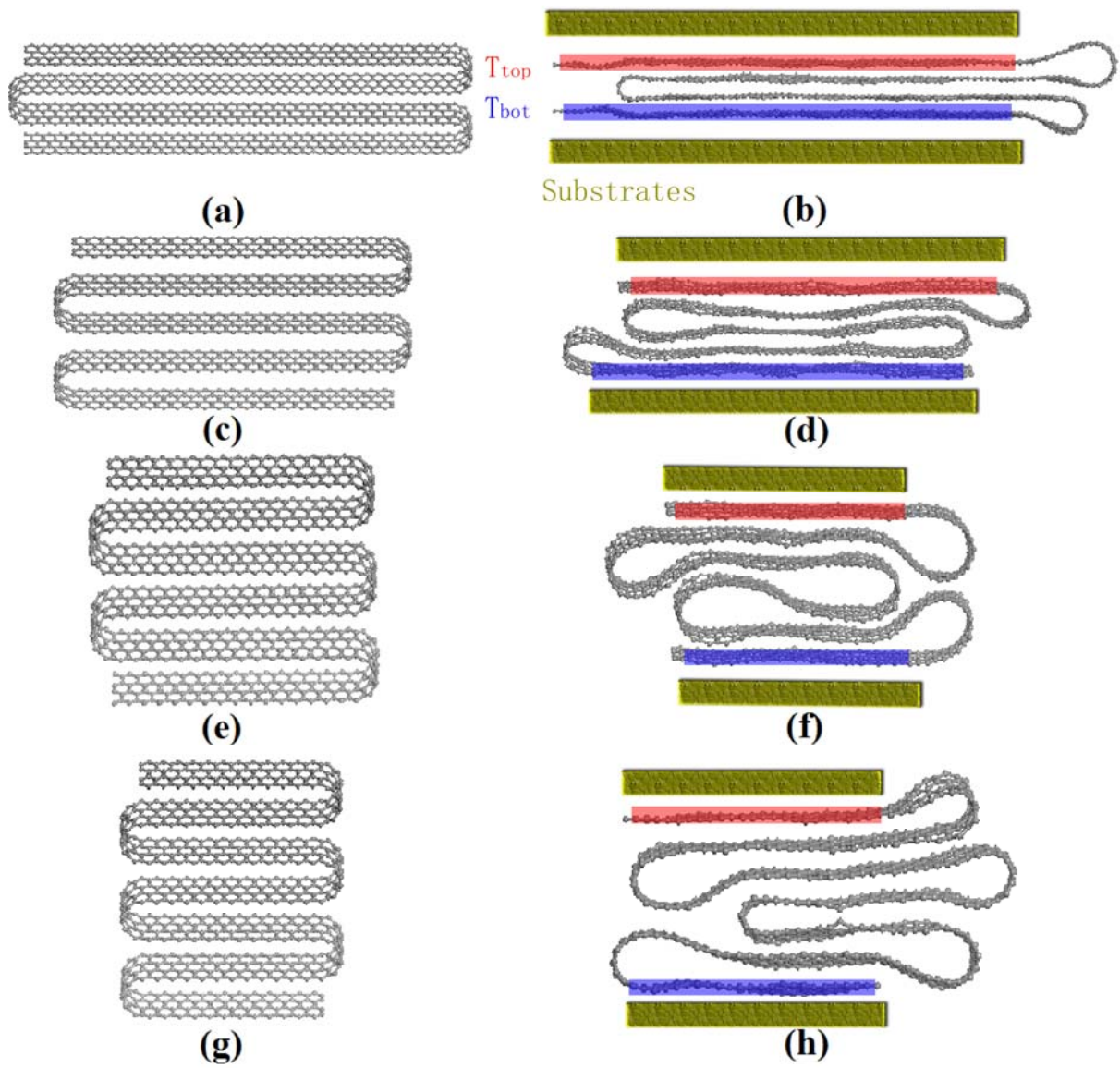

**Fig. 1**

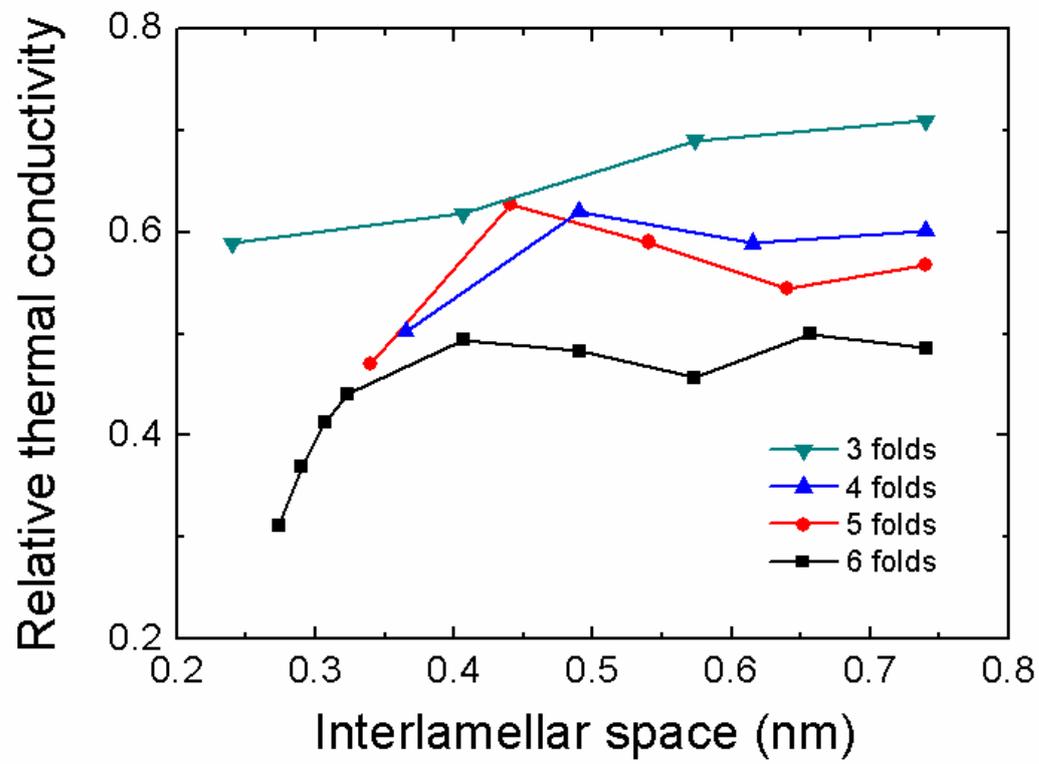

Fig. 2

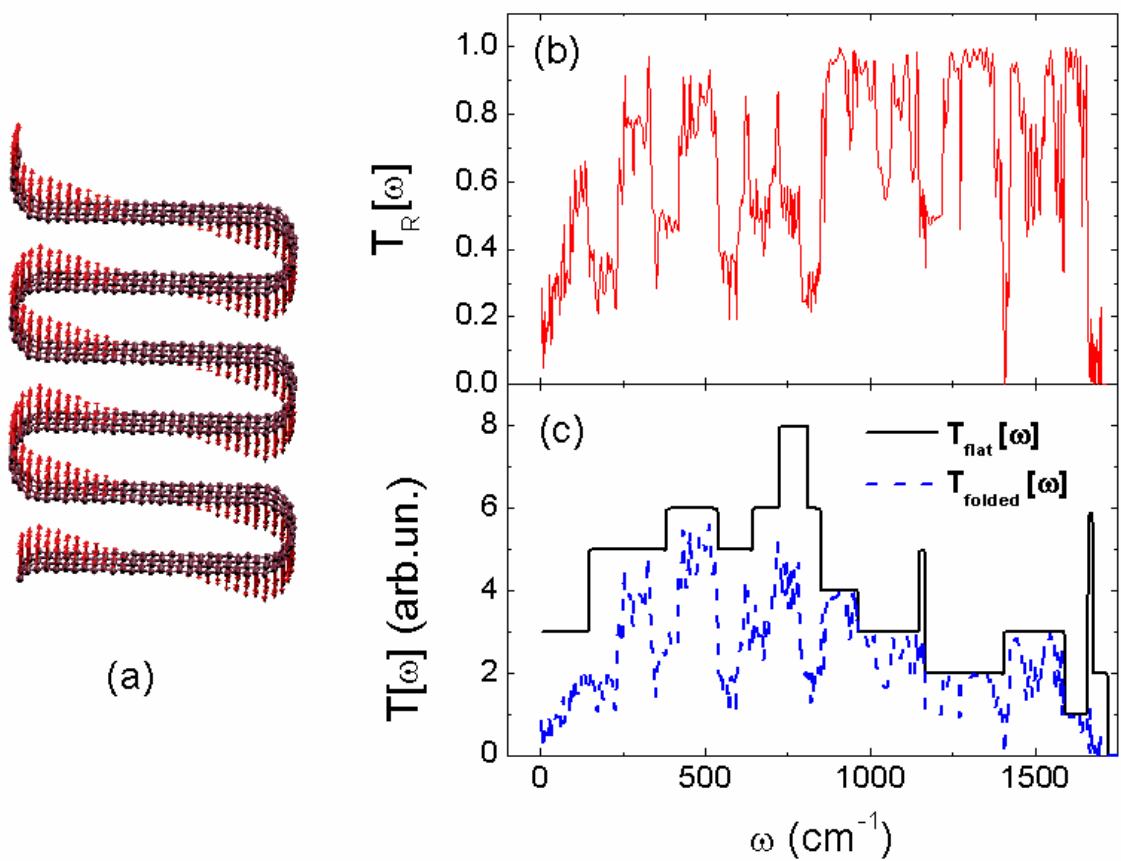

Fig. 3

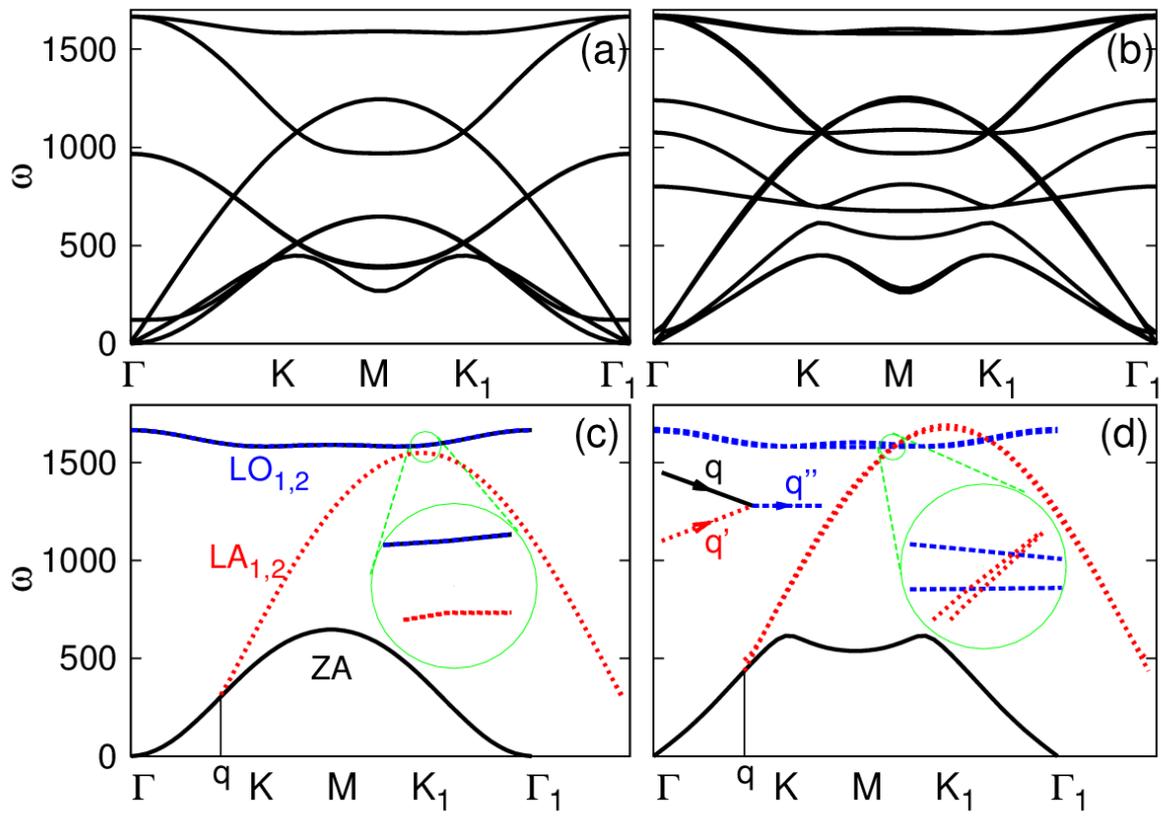

Fig. 4